\documentclass[runningheads]{llncs}
\usepackage{graphicx}
\usepackage{amsmath}
\usepackage{float}
\usepackage[caption=false]{subfig}
\usepackage{tabularx}
\usepackage{multirow}
\usepackage{bm}

\begin{document}

\title{PGT: Pseudo Relevance Feedback Using a Graph-Based Transformer 
}
%
%
\author{HongChien Yu \and
Zhuyun Dai \and
Jamie Callan}

\authorrunning{H. Yu et al.}

\institute{Carnegie Mellon University, Pittsburgh, USA \\
\email{\{hongqiay, zhuyund, callan\}@cs.cmu.edu}
}

\maketitle              
\begin{abstract}
Most research on pseudo relevance feedback (PRF) has been done in vector space and probabilistic retrieval models. This paper shows that Transformer-based rerankers can also benefit from the extra context that PRF provides. It presents PGT, a graph-based Transformer that sparsifies attention between graph nodes to enable PRF while avoiding the high computational complexity of most Transformer architectures. Experiments show that PGT improves upon non-PRF Transformer reranker, and it is at least as accurate as Transformer PRF models that use full attention, but with lower computational costs.

\end{abstract}

\section{Introduction}

Pseudo relevance feedback (PRF) uses context defined by the top-ranked documents of an initial retrieval to improve a subsequent retrieval.  Most prior research has been done in vector space \cite{rocchio}, probabilistic \cite{bm25}, and language modeling \cite{rm12,medmm,dmm} retrieval models. 

Recently the field has moved to Transformer-based rerankers \cite{bertreranker} that are more accurate and computationally complex.  Most Transformer-based rerankers learn contextualized representations from query-document pairs, but they have two limitations. First, the query-document pair provides limited context for query understanding. Second, most Transformers have computational complexity quadratic to the input sequence length, rendering longer context infeasible. 

To overcome these limitations, we propose a PRF method using a graph-based Transformer (PGT). PGT constructs a graph of the query, the candidate document, and the feedback documents. It uses intra-node attention to contextualize the query according to each individual document, and it uses inter-node attention to aggregate information. With the graph approach, PGT can utilize richer relevance context using a configurable number of feedback documents. Its inter-node attention is sparsified, so it also saves computation.  

This paper makes two contributions to the study of pseudo relevance feedback in Transformer architectures.  First, it investigates several ways of using PRF documents as context for Transformer rerankers. It shows that PGT improves upon non-PRF Transformer rerankers, and that PGT is at least as accurate as Transformer PRF models that use full attention, while reducing computation.  Second, it studies the impact of contextual interactions by adjusting the configuration of the graph.  It shows that token-level interaction between the query and feedback documents is critical, while document-level interaction is sufficient to aggregate information from multiple documents.  

\section{Related Work} \label{relatedwork} 

Pseudo-relevance feedback is a well-studied method of generating more effective queries.  Typically pseudo-relevance feedback uses the top-ranked documents to add query terms and set query term weights.  Well-known methods include Rocchio \cite{rocchio}, BM25 expansion \cite{bm25}, relevance models \cite{rm12}, and KL expansion models \cite{medmm,dmm}. A large body of work studies which documents to use for expansion (e.g., \cite{sigir07-kct}).  Most methods were designed for discrete bag-of-words representations.

Recent research also studies PRF in neural networks. Li et al. \cite{nprf} present a neural PRF framework that uses a feed forward network to combine the relevance scores of feedback documents. Only marginal improvement was observed over simple score summation, indicating that the framework does not make the best use of the feedback documents' information. 

Recently, pre-trained Transformer \cite{attention} language models, such as BERT \cite{bert},  have improved the state-of-the-art for ad hoc retrieval. Most Transformer-based rerankers are applied to individual query-document pairs. Some research explores jointly modeling multiple top retrieved documents in a Transformer architecture for question clarification \cite{hashemi2020guided}, question answering \cite{realm,orqa} or code generation~\cite{gemmell2020relevance}.  The effectiveness of using top retrieved documents in Transformer rerankers remains to be studied.
 
While the Transformer-based architectures have achieved state-of-the-art results in multiple natural language tasks \cite{bert}, the original self-attention mechanism incurs computational complexity quadratic to the length of the input sequence. Therefore, much recent work studies sparsifying Transformer attention \cite{longformer,sparse_transformer,transformerxh}. Among these models, Transformer-XH \cite{transformerxh} features an underlying graph structure, where each node represents a text sequence, which makes it a good candidate for multi-sequence tasks such as PRF.  

Transformer-XH employs full-attention within each sequence, but it sparsifies inter-sequence attention. Specifically, for each document sequence $s$, the $l$th layer encoder calculates the intra-sequence, token-level attention by the standard self-attention. Inter-sequence, document-level attentions are calculated using the hidden representations of each sequence's first token \texttt{[CLS]}: 
\begin{equation} 
    \hat{h}_{s, 0}^l = \sum_{s' \in \mathcal{N}(s)} softmax_{s'}(\frac{\hat{q}_{s, 0}^T \cdot \hat{k}_{s', 0}}{\sqrt{d_{k}}}) \cdot \hat{v}_{s', 0} 
\end{equation} 
, where $\mathcal{N}(s)$ are the neighboring document sequences of $s$ in the graph.  This allows the \texttt{[CLS]} token to carry context from other neighboring sequences. Such information is propagated to other tokens in the sequence through the intra-sequence attention in the next layer.  Hence Transformer-XH outputs a condensed representation that contains both the global graph-level information and the local sequence-level information. 

\section{Proposed Method} \label{graphview} 

\begin{figure}[t]
    \centering
    \includegraphics[width=\textwidth]{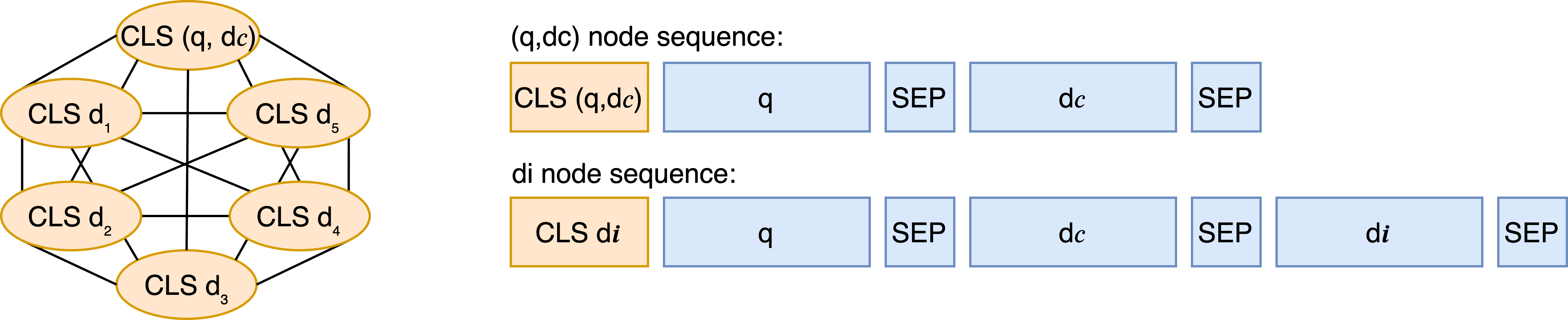}
    \caption{Right: Nodes in PGT contextualize the query using the candidate document $d_c$ and the feedback documents $d_i$ with intra-sequence, token-level attention. Left: The input graph is fully connected with inter-sequence attention among \texttt{[CLS]} tokens. }
    \label{fig:pgt}
\end{figure}

We propose PGT, a PRF reranker with a graph-based Transformer. 
Given a query $q$, a candidate document $d_c$, and feedback documents $d_{1}$, ..., $d_{k}$ retrieved by a first-stage retrieval algorithm, the goal is to predict the score of $d_c$ by aggregating information from feedback documents. To achieve this goal, PGT adopts the Transformer-XH \cite{transformerxh} architecture, and builds a graph of  $q$, $d_c$ and  $d_{1}$, ..., $d_{k}$. Figure \ref{fig:pgt} illustrates the graph. 

PGT has two types of nodes. The $d_i$ nodes contextualize the query using feedback documents. As shown in Fig. \ref{fig:pgt} (right), the input to a $d_i$ node is the text of $d_i$, with $q$ and $d_c$ prepended in order to extract information specific for predicting the relevance between $q$ and $d_c$. The input text sequence is fed into a Transformer module with standard token-level self-attention. To distinguish different parts of the input, we associate segment id 0 with $q$ and $d_c$, and 1 with $d_i$.  In addition to the feedback document nodes, PGT also adopts a special node for the query-candidate pair $(q, d_c)$. The input of the $(q, d_c)$ node is the concatenation of the query and candidate document, which constitutes a typical input sequence to existing Transformer-based rerankers. We hypothesize that the $(q, d_c)$ node will help the model focus more on the query-candidate pair. 

PGT aggregates sequence-level information through inter-sequence attention. Within the sequence, the Transformer encodes the \texttt{[CLS]} token to represent the whole sequence (Fig \ref{fig:pgt} right). Between the sequences, all \texttt{[CLS]} tokens attend to each other to gather information from other sequences (Fig \ref{fig:pgt} left). We follow Zhao et al. \cite{transformerxh} and incorporate inter-sequence attention in the last three Transformer encoder layers. The model is trained on a binary relevance classification task using cross-entropy loss, and it predicts the final relevance score using a weighted sum of all the \texttt{[CLS]} representations \cite{transformerxh}. 

\section{Experimental Setup}
This section describes our datasets, baselines and other experimental settings.  

\subsection{Datasets} 
Experiments were done with the MS MARCO Passage Ranking task dataset \cite{msmarco}.
It contains about 8.8 million passages and about 0.5
million queries with relevance judgments as training data. 
Each query has an equal number of relevant and non-relevant passages. We used the official
evaluation query set from the TREC 2019 Deep Learning Track \cite{trec2019}.
It contains 43 test queries manually annotated by NIST on a four-point scale. On average, a query has 95 relevant documents. 
 We report NDCG@10, MAP@10, and MAP@100. 

\subsection{Baselines} 
We compare PGT to initial rankers, a non-PRF reranker, and PRF models. 
\begin{itemize}
\item{\textbf{BM25 (initial ranker):}} We used Anserini's implementation \cite{anserini}. $k_1$ and $b$ were tuned using a parameter sweep on 500 training queries, following \cite{deepct}. 
\item{\textbf{CRM (initial ranker):}} This model combines BM25's lexical retrieval and BERT's dense embedding retrieval.  It performs significantly better than BM25 on our dataset. We used the rankings provided by Gao, et al.~\cite{clear}.
\item{\textbf{BERT reranker (non-PRF reranker):}} This is a standard BERT reranker, whose input is the concatenated sequence of the query $q$ and the candidate document $d_c$. We trained the model following Nogueira and Cho.~\cite{bertreranker}. 
\item{\textbf{RM3 (PRF):}} This is a traditional language modeling PRF method \cite{trec2004,rm12}.
\item{\textbf{BERT PRF (PRF):}} This is the same as BERT reranker except that we concatenate $(q, d_c, d_{1}, d_{2}, ..., d_{k})$ to form a PRF input sequence, with documents separated by \texttt{[SEP]}. Limited by the input length constraint of BERT~\cite{bert}, we used 5 feedback documents. Same as for PGT, we used segment id 0 for $q$ and $d_c$, and 1 for $d_i$. 
\end{itemize}

\subsection{PGT Graph Variants} \label{graphvariants} 
Modeling queries and documents in a graph gives control over how representations are contextualized. We examined 5 graph variants to study this effect.
\begin{itemize}
    \item{\textbf{PGT base}} is the graph described in Section \ref{graphview}. The query is first contextualized by the candidate and feedback document at the token-level. Feedback information is then aggregated following the graph structure. The $(q, d_c)$ node emphasizes $q$ and $d_c$ at the graph-level. This variant has the richest context. 
    \item{\textbf{PGT w/o pre} $\bm{d_c}$} removes prepended candidate from the $d_i$ nodes, so each query is only  contextualized by the feedback document at sequence-level. 
    \item{\textbf{PGT w/o pre} $\bm{q, d_c}$} removes both the prepended query and the prepended candidate from the feedback nodes. Each feedback document hence only contextualizes the query at the graph-level.
    \item{\textbf{PGT w/o node} $\bm{d_c}$} removes candidate from the $(q, d_c)$ node, so only $q$ is emphasized again at the graph-level.
    \item{\textbf{PGT w/o node} $\bm{q, d_c}$} removes the $(q, d_c)$ node from the graph, so $q$ and $d_c$ are not emphasized again at the graph-level. 
\end{itemize}

\subsection{Training and Evaluation} \label{depth}

We implement PGT  based on the Transformer-XH~\cite{transformerxh} PyTorch implementation. The parameters for the intra-sequence attention are initialized from a pre-trained BERT base model \cite{bert}, and those for the inter-sequence attention are initialized according to Xavier et al. \cite{xavier}. We train the model for 2 epochs, with per-GPU batch size = 4 on 2 GPUs. The maximum node sequence length is 128, and the learning rate is 5e-6 with linear decay.  

We train both BERT PRF and PGT using feedback documents from BM25. In order to test how Transformer-based PRF models generalize when different initial rankers are used, we evaluate them using both BM25 and CRM. 
We follow prior research \cite{clear,bertreranker} and report the results at each model's best reranking depth $r$ (Table \ref{tab:main}).  

\section{Experimental Results}
\textbf{PRF vs. non-PRF Transformers} We study the effectiveness of PRF in Transformer-based models by comparing PGT and BERT PRF with BERT reranker.  Table \ref{tab:main} shows that all PRF Transformers outperform BERT reranker on MAP@10 using either initial ranker. In particular, PGT achieves MAP@10 $13.0\%$ and $7.4\%$ better than BERT reranker on BM25 and CRM respectively, with comparable NDCG@10. The results suggest that the richer context provided by PRF helps Transformers rank relevant documents to the very top. 

PRF enables Transformers to exploit high-quality initial rankings better. Comparing BM25 and CRM results in Table \ref{tab:main}, we found that when the initial ranker is stronger, PGT achieves the best performance across all metrics, closely followed by BERT PRF. In comparison, BERT reranker cannot make the best of the initial retrieval of CRM, as reported by prior research \cite{clear}.  

\textbf{PGT vs. BERT PRF} While PGT rankings are at least as good as BERT PRF, it is more computationally efficient. Using $k=5$ for a fair comparison, we calculated the number of multiplication and addition operations. PGT consumes $88\%$ as many operations on each input example compared with BERT PRF. In addition, PGT requires smaller reranking depth (Table \ref{tab:main}). Using BM25 as the initial ranker, the computational cost is hence only $44\%$ of BERT PRF's.   

Compared with BERT PRF, PGT allows flexible configurations on the graph structure (Table \ref{tab:main}). As discussed in Section \ref{graphvariants}, the graph structure controls how relevance context flows across the graph. Contrary to our initial intuition, removing the $(q, d_c)$ node partially or entirely (\text{PGT w/o node $d_c$} and \text{PGT w/o node $q, d_c$}) achieves the best results among all graph variants. $q$ is an impoverished description of the information need compared to feedback documents $d_1 \ldots d_k$, which may explain why the comparison of $q$ to $d_c$ is less useful than comparisons between $d_c$ and high-quality documents.

The number of feedback documents $k$ is a parameter that is usually tuned.  BERT's self-attention mechanism restricts the input sequence length, limiting BERT to 5 feedback documents on our dataset.  PGT has no such restriction.  Our experiments use $k=7$ for PGT because it is more effective (Table \ref{tab:pgtk}).

\begin{table}[t] 
\centering
\caption{The evaluation results with BM25 and CRM as initial rankers. RM3 is shown for completeness, but it is not competitive, so it is not discussed. We report the results at each models' best reranking depth ($r$) according to prior research \cite{clear,bertreranker}. We use $k=7$ feedback documents for PGT. $*$ and $\dagger$ indicate statistical significance over the initial ranker and BERT reranker using t-test with $p \leq 0.05$. } 

\label{tab:main}
\begin{tabular}{|c|cccc|cccc|}
\hline
               & \multicolumn{4}{c|}{\textbf{ BM25}}        & \multicolumn{4}{c|}{\textbf{CRM}} \\
               & NDCG & MAP    & MAP      &                       & NDCG   & MAP       & MAP &          \\
               & @10 & @10    & @100      & r                     & @10    & @10       & @100    & r           \\ \hline \hline
Initial Ranker & 0.5058  & 0.1126 & 0.2993  & -                    & 0.6990  & 0.1598    & 0.4181   & -      \\
RM3            & 0.5180  & 0.1192 & 0.3370$^{*}$  & 1K         & \multicolumn{4}{c|}{-- \footnotemark{}}   \\
BERT Reranker  & 0.6988$^{*}$  & 0.1457$^{*}$ & 0.3905$^{*}$   & 1K    & 0.7127  & 0.1572  & 0.4134 & 20       \\
 \hline \hline
BERT PRF       & 0.6862$^{*}$  & 0.1495$^{*}$ & \textbf{0.4075}$^{*}$  & 1K     & 0.7188    & 0.1646   & 0.4203  & 20      \\
PGT base      & 0.6712$^{*}$  & 0.1542$^{*}$ & 0.3927$^{*}$ & 500     & 0.7238$^{*}$  & 0.1660 & 0.4205  & 20 \\ 
\multicolumn{1}{|l|}{PGT \text{w/o pre $d_c$} }     & 0.6693$^{*}$  & 0.1523$^{*}$  &  0.3563$^{*}$  & 500   & 0.7146  &  0.1658   & 0.4194  & 20         \\
\multicolumn{1}{|l|}{PGT \text{w/o pre $q, d_c$}  }     & 0.6676$^{*}$  & 0.1468$^{*}$  &  0.3450$^{*}$  & 500   & 0.7005  &  0.1572   & 0.4145  & 20      \\
\multicolumn{1}{|l|}{PGT \text{w/o node $d_c$}  }    & 0.6840$^{*}$  & 0.1586$^{*}$  &  0.3868$^{*}$  & 500                  & 0.7139  &  \textbf{0.1689}$^{*}$   & 0.4192  & 20      \\
\multicolumn{1}{|l|}{PGT \text{w/o node $q, d_c$} }      & \textbf{0.7078}$^{*}$  & \textbf{0.1646}$^{*\dagger}$  &  0.3819$^{*}$  & 500 & \textbf{0.7326}$^{*}$  & 0.1654   & \textbf{0.4220}  & 20  \\ \hline
\end{tabular}
\end{table}

\footnotetext{CRM jointly trains a hybrid of sparse and dense retrieval models. Running RM3 on CRM is an open question that is beyond the scope of this work.}

\begin{table}[t]
\centering
\caption{ PGT base using different numbers of feedback documents $(k)$ }
\label{tab:pgtk} 
\begin{tabular}{|c|ccc|ccc|}
\hline
  & \multicolumn{3}{c|}{\textbf{BM25}} & \multicolumn{3}{c|}{\textbf{CRM}}           \\
  & NDCG  & MAP    & MAP                & NDCG   & MAP    & MAP     \\
k & @10  & @10    & @100                & @10    & @10    & @100    \\ \hline \hline 
5 & 0.6344 & 0.1497 & 0.3536            & 0.6923 & 0.1653 & 0.4177 \\
7 & \textbf{0.6712} & \textbf{0.1542}  & 0.3927 & \textbf{0.7238}  & \textbf{0.1660} & \textbf{0.4205} \\
9 & 0.6538  & 0.1476 & \textbf{0.3931}  & 0.6940 & 0.1636 & 0.4180 \\ \hline
\end{tabular}
\end{table}

\section{Conclusion}
Most Transformer-based rerankers learn contextualized representations for query-document pairs, however queries are impoverished descriptions of information needs.  This paper presents PGT, a pseudo relevance feedback method that uses a graph-based Transformer.  PGT graphs treat feedback documents as additional context and leverage sparse attention to reduce computation, enabling them to use more feedback documents than is practical with BERT-based rerankers.

Experiments show that PGT improves upon non-PRF BERT rerankers. Experiments also show that PGT rankings are at least as good as BERT PRF rerankings, however they are produced more efficiently due to fewer computations per document and fewer documents reranked per query. PGT is robust, delivering effective rankings under varied graph structures and with two rather different initial rankers.

%
%
%
\bibliographystyle{splncs04}
\bibliography{mybibliography}
\end{document}